\documentstyle[12pt]{article}
\begin{document}
\baselineskip=24pt

\begin{center}{\bf Fermion-Antifermion Condensate Contribution to the
Anomalous Magnetic Moment of a Fundamental Dirac Fermion}
\end{center}

\bigskip

\bigskip

\begin{center}Victor Elias \footnote{Corresponding Author - \\
\indent tel:  519-679-2111, ext 8785 \\
\indent fax:  519-661-3523 \\
\indent email: zohar@apmaths.uwo.ca} and Kevin Sprague \\
Department of Applied Mathematics \\
The University of Western Ontario \\
London, Ontario  N6A 5B7  CANADA
\end{center}

\vfill

\noindent Running Head:  {\it Condensate Contribution to the Fermion
Magnetic Moment}

\bigskip

\newpage

\begin{center}
{\it Abstract}
\end{center}

We consider the contribution of fermion-antifermion condensates to
the anomalous magnetic moment of a fermion in a vacuum in which such
condensates exist.  The real part of the condensate contribution to
the anomalous magnetic moment is shown to be zero.  A nonzero
imaginary part is obtained below the kinematic threshold for
intermediate fermion-antifermion pairs.  The calculation is shown to
be gauge-parameter independent provided a single fermion mass
characterizes both the fermion propagator and condensate-sensitive
contributions, suggestive of a dynamically-generated fermion mass. 
The nonzero imaginary part is then argued to correspond to the
kinematic production of the intermediate-state Goldstone bosons
anticipated from a chiral-noninvariant vacuum.  Finally, speculations
are presented concerning the applicability of these results to quark
electromagnetic properties.

\vfill
 

\newpage 

\noindent{\bf 1.  INTRODUCTION}

One of the key distinctions between quantum electrodynamics (QED) and quantum
chromodynamics (QCD), field-theories of known-interaction physics with unbroken gauge
symmetry, is the existence of QCD-vacuum condensates, a distinction that has not been
adequately linked to the nonabelian character of the latter theory.  In particular, 
the quark-antiquark condensate $<\bar{q}q>$ characterizes the chiral noninvariance of the QCD
vacuum;
QED has no corresponding electron-positron condensate.  Although a dynamical breakdown in
chiral invariance can be linked to a criticality-threshold in the size of the gauge coupling constant
(Higashijima, 1984), such an argument might in-and-of itself suggest the possible existence of
an electron-positron condensate in the electromagnetic potential of atomic nuclei with sufficiently
large atomic number.  In such a scenario, the only linkage between the nonabelian character of
QCD and the existence of vacuum-condensates would be the large size of the running QCD
coupling-constant anticipated in the low-momentum- (static-quark-) limit. 

It is therefore of interest to explore whether the existence of such a condensate might alter 
well-understood static fermion properties of QED.  In QCD, particularly QCD sum-rule
applications
(Shifman {\it et al.,} 1979), such condensates characterize the vacuum expectation values of
normal ordered products of fields (Pascual and Tarrach, 1984). For example (Elias {\it et al.,}
1988; Yndurain, 1989), the quark-antiquark condensate characterizes the QCD-vacuum
expectation-value 

\renewcommand{\theequation} {1.1\alph{equation}}
\setcounter{equation}{0}

\begin{equation}
<0|:\psi_i^a (y) \bar{\psi}_j^b (z):|0> = -\frac{\delta}{3}^{ab}
<\bar{q}q> \sum_{r=0}^\infty C_r \left[ -im \gamma^\mu (y_\mu -
z_\mu) \right]_{ij}^r ,
\end{equation}
\begin{equation}
C_r = \left\{ \begin{array}{c}\frac{1}{(r/2)![(r+2)/2]!2^{r+2}}\; \; , \; \;
r \; {\rm even}, \\
\frac{1}{[(r - 1)/2]! [(r+3)/2]! 2^{r+2}} \; \; , \; \; r \; {\rm
odd}.
\end{array}
\right\}
\end{equation}
[{i,j} are Dirac indices and {a,b} are colour indices.] In principle, every Feynman amplitude
whose Wick-Dyson expansion of time-ordered fields contains a term in which a fermion and
antifermion field are contracted to form a propagator,

\renewcommand{\theequation}{1.\arabic{equation}}
\setcounter{equation}{1}

\begin{equation}
<0|T \psi_i^a (y) \bar{\psi}_j^b
(z)|0> = \int \frac{d^4 p}{(2 \pi)^4} e^{-i p \cdot (y - z)} \left[
S_F (p) \right]_{ij}^{ab} , 
\end{equation}
also contains a not-fully contracted term containing (1.1).  It is precisely through such terms,
methodologically, that QCD-vacuum condensates are introduced into the field-theoretical side of
sum-rule calculations (Pascual and Tarrach, 1984). It should be noted that the
{\it only} signature of
nonabelian physics in the derivation of (1.1) is an overall colour-summation factor of 
$3 (= \delta^{aa})$ in the denominator, a factor that can easily be absorbed in a 
redefinition of the fermion-antifermion condensate $<\bar{f}f>$ for the abelian case. [We will
define
$<\bar{f}f>$ henceforth such that $\delta^{ab}<\bar{q}q> \; \; \rightarrow
\; \; <\bar{f}f>$ in (1.1a).] The result (1.1) can be otherwise understood as a solution to the free
Dirac
equation with the condensate entering through an appropriately chosen initial condition
(Yndurain, 1989). This nonzero condensate is a reflection of the nonperturbative content of the
vacuum.  Normal-ordered fields necessarily annihilate a purely-perturbative vacuum, which is
why vacuum expectation values like the left-hand side of (1.1) are
{\it not} incorporated into standard
QED calculations, but {\it are} incorporated into the field-theoretical content of QCD sum-rules.

      In this paper, we address whether the explicit fermion-antifermion condensate contribution
to the anomalous QED magnetic moment of a fermion field is calculable
through use of field-theoretical
techniques for this contribution (Bagan {\it et al.,} 1993 and 1994) adapted from QCD sum-rule
applications. We are motivated to examine this particular property because it can be easily
extracted from the leading corrections to the unrenormalized electromagnetic vertex-function
without reference to self-energy, vacuum polarization, or bremsstrahlung graphs that enter into
the determination of electromagnetic form-factor slopes. 

In Section 2 of this paper, we provide a brief methodological review of how the
anomalous magnetic moment ${\cal K} F_2(0)$ of a Dirac fermion 
with mass $m$ is calculated in (purely-) perturbative
QED. In Section 3, we modify this calculation, as indicated above, by including the contribution
of vacuum expectation values (1.1) in the Wick-Dyson expansion of the unrenormalized QED
vertex amplitude. We find that the real part of the 
$<\bar{f}f>$-contribution to ${\cal K} F_2(0)$ vanishes, but that
an imaginary part develops for $q^2$ between zero and $4m^2$ which diverges 
in the $q^2 \rightarrow  0$ limit.

These results are discussed in Section 4.  They are first shown to be gauge parameter independent
provided the same fermion mass characterizes (1.1) and (1.2), results suggestive of a dynamical
rather than a Lagrangian origin for the common fermion mass.
We then argue that a change in the kinematic threshold for the production of
physical elementary particle states is the most sensible interpretation of the imaginary part
obtained in Section 3, indicative of the production of Goldstone
bosons anticipated from the dynamical breakdown of chiral symmetry
($<\bar{f}f> \neq 0$). 

Up to this point in the paper, the question of condensate contributions to the anomalous
magnetic moment has been posed entirely in the abstract.  In Section 4, we discuss the
applicability of the results of Section 3 to quarks, as fundamental Dirac fermions which form
nonzero $<\bar{q}q>$ condensates whose contribution to ${\cal K}F_2(0)$ 
is precisely of the type investigated in
Section 3.  Although quarks are confined, their QED magnetic moments and form-factors are
nevertheless of phenomenological interest for extracting baryon magnetic moments (Beg
{\it et al.,} 1964, and Perkins, 1987) and form-factor behaviour. 
In the {\it absence} of condensate contributions,
${\cal K} þF_2(q^2)$ develops an imaginary
part when $q^2 > 4m^2$, corresponding to the production of on-shell 
fermion-antifermion pairs.  $<\bar{q}q>$
contributions are seen to reduce this threshold to $q^2 = 0$, which 
can most easily be understood
(assuming $m$ is dynamical) to correspond to the kinematical production 
of massless pions, the Goldstone bosons of the chiral symmetry breaking 
whose order-parameter manifestation is the
$<\bar{q}q>$ condensate itself.  Thus, the change in the onset of an 
imaginary part in 
${\cal K} F_2(q^2)$ may reflect
the transition of QCD to low-energy hadronic physics. 

\bigskip

\noindent{\bf 2. THE ELECTROMAGNETIC VERTEX CORRECTION:  A
METHODOLOGICAL REVIEW}

The purely-perturbative three-point Green's function [Fig. 1]
containing the truncated fermion-antifermion-photon vertex Green's
function $-i Q e \Gamma^\sigma (p_2, p_1)$ is expressed in terms of
Heisenberg fields $\psi_i, \bar{\psi}_j, A_\mu$ as
follows:

\renewcommand{\theequation}{2.\arabic{equation}}
\setcounter{equation}{0}
\begin{eqnarray}
& &[G_\mu (p_2, p_1)]_{i\ell} \nonumber \\
& = & \left[ \frac{-i}{(p_2 - p_1)^2} \left(
g_{\mu\sigma} - (1 - \xi) \frac{(p_2-p_1)_\mu
(p_2-p_1)_{\sigma}}{(p_2-p_1)^2} \right) \right] \left[
\frac{i}{\not{p}_2 - m} \right]_{ij} \nonumber \\
& \cdot & \left[ \frac{i}{\not{p}_1 - m} \right]_{k\ell} \left[ -i e
Q \Gamma_{j k}^\sigma (p_2, \; p_1) \right] , \nonumber \\
& = & \int d^4 x' \int d^4 y' <0|T \psi_i (x') A_\mu (0)
\bar{\psi}_{\ell}
(y')|0>_{\small Heis} e^{+i p_2 \cdot x'} e^{-i p_1 \cdot y'} ,
\end{eqnarray}
where $i-\ell$ are Dirac indices.  The unrenormalized one-loop vertex
correction $\Lambda^\mu$ within $\Gamma^\mu$ $(= \gamma^\mu +
\Lambda^\mu + ...)$ is obtained via a Wick-Dyson expansion of the
vacuum expectation value in (2.1) evaluated in the interaction
picture (Dirac indices have been dropped):
\begin{eqnarray}
<0|T \psi (x') A_\mu (0) \bar{\psi} (y')|0>_{\small Heis} 
& = &<0|T \psi (x') \exp \left[ -iQe \int d^4 w \;  \bar{\psi}(w)
\gamma^\tau \psi (w) A_\tau(w) \right]\nonumber \\
& \times & A_\mu (0)\bar{\psi}
(y')|0>,
\end{eqnarray}
The one-loop correction to $G_\mu$ in (2.1) is then found to be [Fig.
2]
\begin{eqnarray}
& & \left[ \Delta  G_\mu (p_2, p_1) \right]^{1-loop}
\nonumber \\
& = & (-i Q e)^3 \int d^4 x' \int d^4 y' e^{+i p_2 \cdot x'} e^{-i
p_1 \times y'} \nonumber \\
& \times & \left[ \int d^4 x \int d^4 y \int d^4 z <0|T \psi(x')
\bar{\psi}(x)|0> \gamma^\tau \right. \nonumber \\
& \times & <0|T \psi(x) \bar{\psi}(y)|0> \gamma^\sigma <0| T \psi(y)
\bar{\psi} (z)|0> \gamma^\rho \nonumber \\
& \times & <0|T \psi(z) \bar{\psi} (y')|0> <0|T A_\mu (0) A_\sigma
(y)|0> \nonumber \\
& \times & \left. <0|T A_\tau (x) A_\rho (z)|0> \right],
\end{eqnarray}
where the term in large square brackets is just the fully-contracted
third-order term in the Wick-Dyson expansion of (2.2).  Eq. (2.3) can be 
evaluated by explicit use of the configuration-space fermion and photon
propagators
\begin{equation}
<0|T \psi (x) \bar{\psi}(y)|0> = i \int \frac{d^4 q}{(2 \pi)^4} e^{-i
q \cdot (x-y)} \frac{1}{\not{q} - m},
\end{equation}
\begin{equation}
<0|T A_\tau (x) A_\rho (z)|0> = -i \int \frac{d^4 k}{(2 \pi)^4} e^{-i
k \cdot (x-z)} \frac{g_{\tau\rho}}{k^2}.
\end{equation}
We have omitted Dirac and colour indices from (2.4), as well as the
gauge-dependent longitudinal term from (2.5), as
it does not contribute to the vertex function.  Upon substitution of
configuration-space propagators (2.4,5) into (2.3) and integration
over the configuration space variables $\{x', y', x, y, z\}$, one
obtains a string of delta functions which, when integrated over,
yield the usual momentum-space Feynman propagator functions:

\newpage

\begin{eqnarray}
& &\left[ \Delta G_\mu (p_2, p_1) \right]_{i \ell}^{1-loop} \nonumber \\
& & = (-i Qe)^3 \int \frac{d^4 q_1}{(2 \pi)^4} \int \frac{d^4 q_2}{(2
\pi)^4} \int \frac{d^4 q_3}{(2 \pi)^4} \int \frac{d^4q_4}{(2 \pi)^4}
\int \frac{d^4 k_1}{(2 \pi)^4} \int \frac{d^4 k_2}{(2 \pi)^4} \nonumber \\
& \times & \left( \frac{i}{\not{q}_1 - m} \right) \gamma^\tau \left(
\frac{i}{\not{q}_2 - m} \right) \gamma^\sigma \left(
\frac{i}{\not{q}_3 - m} \right) \gamma^\rho \left( \frac{i}{\not{q}_4
- m} \right) \nonumber \\
& \times & \left( - \frac{i g_{\mu \sigma}}{k_1^2} \right) \left( - \frac{i
g_{\tau \rho}}{k_2^2} \right) (2 \pi)^{20} \delta^4 (q_1 - p_2)
\delta^4 (q_4 - p_1) \nonumber \\
& \times & \delta^4 (p_2 - q_2 - k_2) \delta^4 (q_2 - q_3 + k_1) \delta^4 (q_3 -
p_1 + k_2) \nonumber \\
& = & \left[ - \frac{i g_{\mu \sigma}}{(p_2 - p_1)^2} \right] \left[
\frac{i}{\not{p}_2 - m} \right]_{ij} \left[ \frac{i}{\not{p}_1 - m}
\right]_{k \ell} (-i Qe) \nonumber \\ 
& \times & \left\{ - \frac{i(Qe)^2}{(2 \pi)^4} \int \frac{d^4
k_2}{k_2^2} \gamma^\tau \frac{ (\not{p}_2 - \not{k}_2 + m)}{(p_2 -
k_2)^2 - m^2} \gamma^\sigma \frac{(\not{p}_1 - \not{k}_2 + m)}{(p_1 -
k_2)^2 - m^2} \gamma_\tau \right\}_{j k}
\end{eqnarray}

Factorization of the external legs is explicit in the final line of
(2.6).  The curly bracketed expression in the final line of (2.6)
corresponds to the unrenormalized vertex correction $[\Lambda^\sigma
(p_2, p_1)]$.  Specifically, one can define the unrenormalized vertex
correction to be $\bar{u} (p_2) \Lambda^\mu (p_2, p_1) u (p_1)$, where
$[q^\mu \equiv p_2^\mu - p_1^\mu]$
\begin{equation}
\Lambda^\mu (p_2, p_1) \equiv e^2 Q^2 \left[ R(q^2) \gamma^\mu +
\frac{2 S(q^2)}{m} (p_1^\mu + p_2^\mu) \right],
\end{equation}
such that the unrenormalized vertex is $-i e Q \Gamma^\mu \equiv -i e
Q \left( \gamma^\mu + \Lambda^\mu (p_2, p_1) \right)$ with $e Q$ the
electromagnetic fermion charge.  This unrenormalized vertex can be
expressed as follows in terms of the renormalized vertex form 
factors $F_1(q^2), \; \; {\cal K} F_2(q^2)$:
\begin{eqnarray}
& &\bar{u} (p_2) \left[ \gamma^\mu + \Lambda^\mu (p_2, p_1) \right]
u(p_1) \nonumber \\
& = & \bar{u} (p_2) \left[ ( 1 + e^2 Q^2 [R(q^2) + 4S(q^2) ]) \gamma^\mu
- 2 e^2 Q^2 S(q^2) i \sigma^{\mu\nu} q_\nu / m \right] u(p_1)
\nonumber \\
& \equiv & Z \; \bar{u} (p_2) \left[ F_1 (q^2) \gamma^\mu + i
\sigma^{\mu\nu} q_\nu {\cal K} F_2 (q^2) / 2m \right] u(p_1) .
\end{eqnarray}
The rescaling in the final line of (2.8) is accomplished through the
renormalization condition that $F_1(0) = 1$, in which case the
(divergent) constant $Z$ is given to order-$e^2$ by
\begin{equation}
Z = 1 + e^2 Q^2 (R(0) + 4S(0)) .
\end{equation}
To leading order in $e^2$, one then finds that
\begin{equation}
F_1 ' (q^2) = 1 + e^2 Q^2 \left[ (R'(0) + 4S'(0) ) q^2 + {\cal O}(q^4)
\right] + {\cal O}(e^4),
\end{equation}
\begin{equation}
{\cal K} F_2 (q^2) = -4e^2 Q^2 S(q^2) + {\cal O}(e^4) .
\end{equation}

The $q^2 \rightarrow 0$ limit of Eq. (2.11) gives the ${\cal
O}(\alpha)$ anomalous magnetic moment of QED, which (in contrast to
$F_1$) devolves solely from the vertex correction and is insensitive
to additional (vacuum-polarization, self-energy, and bremsstrahlung)
diagrams.

The purely perturbative contribution to this quantity can be extracted
by straightforward methods from the unrenormalized vertex correction
in (2.6):
\begin{eqnarray}
\left[ \Lambda^\mu (p_2, p_1) \right]^{pert} & = & -i (Qe)^2 \int
\frac{d^4 k}{(2 \pi)^4} \frac{\gamma_\tau (\not{p}_2 - \not{k} + m)
\gamma^\mu (\not{p}_1 - \not{k} + m)\gamma^\tau}{k^2 [(p_2 - k)^2 -
m^2][(p_1 - k)^2 - m^2]} \nonumber \\
& = & -i (Qe)^2 \left\{ \left[ -2 \not{p}_1 \gamma^\mu \not{p}_2 + 4m
(p_1^\mu + p_2^\mu) - 2m^2 \gamma^\mu \right] I(p_2, \; p_1) \right.
\nonumber \\
& + & \left[ 2 \gamma^\rho \gamma^\mu \not{p}_2 + 2 \not{p}_1
\gamma^\mu \gamma^\rho - 8mg^{\mu\rho} \right] I_\rho (p_2, \; p_1)
\nonumber \\
& - & \left. 2 \gamma^\rho \gamma^\mu \gamma^\sigma I_{\rho\sigma} (p_2, \;
p_1) \right\}.
\end{eqnarray}

The integrals $I$, $I_\rho$ and $I_{\rho\sigma}$ are respectively
defined by
\begin{eqnarray}
& & \left[ I(p_2, \; p_1) ; \; I_\rho (p_2, \; p_1) ; \;
I_{\rho\sigma} (p_2, \; p_1) \right] \nonumber \\
& \equiv & \int \frac{d^n k}{(2 \pi)^n} \frac{\left[ 1 ; \; k_\rho ;
\; k_\rho k_\sigma \right]}{(k^2-\epsilon^2) \left[(k-p_2)^2 - m^2
\right] \left[ (k-p_1)^2 - m^2 \right]} .
\end{eqnarray}
In (2.13), ultraviolet divergences are regulated via dimensional
regularization and infrared divergences are regulated via the
``photon-mass'' $\epsilon$.  These integrals are easily evaluated by
standard methods.  If we evaluate these integrals in terms of a set
of constants $a_{0,1}, \; b_{0,1}, \; ...$ such that
\begin{equation}
I(p_2, \; p_1) = i \left\{ a_0 / m^2 + a_1 q^2/m^4 + {\cal O} (q^4)
\right\},
\end{equation}
\begin{equation}
I_\rho (p_2, \; p_1) = i \left[ (p_{1 \rho} + p_{2 \rho}) / m^2
\right] \left\{ b_0 + b_1 q^2 / m^2 + {\cal O} (q^4) \right\},
\end{equation}
\begin{eqnarray}
I_{\rho\sigma} (p_2, p_1) &=& i g_{\rho\sigma} \left\{ c_0 + c_1
q^2/m^2 + {\cal O} (q^4) \right\} \nonumber \\
& + & i \left[ \left( p_{1\rho} p_{1\sigma} + p_{2\rho} p_{2 \sigma}
\right) / m^2 \right] \left\{ d_0 + d_1 q^2/m^2 + {\cal O} (q^4)
\right\} \nonumber \\
& + & i \left[ \left( p_{1 \rho} p_{2\sigma} + p_{2\rho} p_{1\sigma}
\right)/m^2 \right] \left\{ e_0+e_1 q^2/m^2+{\cal O}(q^4)
\right\},
\end{eqnarray} 
we can then express the vertex correction (2.12) in the form (2.7):
\begin{eqnarray}
R(q^2) & = & \left[ 4a_0 - 16b_0 + 4c_0 + 4d_0 + 4e_0 \right]
\nonumber \\
& + & (q^2/m^2) \left[ -2a_0 + 4a_1 + 4b_0 - 16b_1 + 4c_1 \right. \nonumber
\\
& + & 4\left. d_1 - 2e_0 + 4e_1 \right] + {\cal O} (q^4/m^4),
\end{eqnarray}
\begin{eqnarray}
S(q^2) & = & \left[ 2b_0 - 2d_0 - 2e_0 \right] \nonumber \\
& + & (q^2/m^2) \left[ 2b_1 - 2d_1 - 2e_1 \right] + {\cal O}
(q^4/m^4) .
\end{eqnarray}
The set of constants are determined via explicit evaluation of
the integrals in (2.13):
\begin{equation}
a_0 = \frac{1}{32 \pi^2} \ell n \left( \frac{\epsilon^2}{m^2} \right)
, \; a_1 = \frac{1}{192 \pi^2} \left[ 1 + \ell n \left(
\frac{\epsilon^2}{m^2} \right) \right];
\end{equation}
\begin{equation}
b_0 = -\frac{1}{32 \pi^2} , \; b_1 = -\frac{1}{192 \pi^2};
\end{equation}
\begin{equation}
c_0 = \frac{1}{64 \pi^2} \left[ -\frac{2}{n-4} - \gamma_E - \ell n
\left( \frac{m^2}{4 \pi \mu^2} \right) + \frac{1}{2} \right], \; c_1
= \frac{1}{384 \pi^2};
\end{equation}
\begin{equation}
d_0 = -\frac{1}{96 \pi^2} , \; d_1 = -\frac{1}{640 \pi^2} ;
\end{equation}
\begin{equation}
e_0 = -\frac{1}{192 \pi^2} , \; e_1 = -\frac{1}{960 \pi^2} .
\end{equation}
As is evident from comparison of (2.11) and (2.18), the anomalous
magnetic moment is clearly finite,
\begin{equation}
{\cal K} F_2 (0) = -8e^2 Q^2 \left( b_0 - d_0 - e_0 \right) = e^2
Q^2/8\pi^2,
\end{equation}
the famous result of Schwinger and Feynman (Schwinger, 1948, and
Feynman, 1949).  The $F_1$
form-factor slope in the vertex correction (2.10) is also a 
classical result of perturbative QED,
\begin{eqnarray}
F_1'(0) & = & e^2 Q^2 \left[ R'(0) + 4S'(0) \right] \nonumber \\
& = & \frac{e^2Q^2}{m^2} \left[ -2a_0 + 4a_1 + 4b_0 - 8b_1 + 4c_1 -
4d_1 - 2e_0 - 4e_1 \right] \nonumber \\
& = & -\frac{e^2 Q^2}{24 \pi^2 m^2} \left[ \frac{3}{4} + \ell n
\frac{\epsilon^2}{\mu^2} \right],
\end{eqnarray}
although the removal of the photon mass from the physically
measurable form-factor slope entails careful consideration of soft
external-photon (bremsstrahlung) diagrams (Block and Nordsieck, 1937;
Yennie {\it et al.,} 1955).

\bigskip

\noindent{\bf 3. ${\bf <\bar{f}f>}$ CONTRIBUTION TO ${\bf {\cal K} F_2
(q^2)}$}

\renewcommand{\theequation}{3.1\alph{equation}}
\setcounter{equation}{0}

In Figure 3a (3b), we replace the perturbative configuration-space
propagator $<0|T \psi (x) \bar{\psi} (y)|0> \; \; (<0|T \psi(y)
\bar{\psi}(z)|0>)$ of Eq. (2.3) with the ``non-perturbative
propagator'' (Yndurain, 1989) $<0|: \psi(x) \bar{\psi}(y):|0>$ $(<0|:
\psi(y) \bar{\psi}(z):|0>)$, the vacuum-expectation value of the
normal-ordered pair of fermion fields that {\it also} arises from the
Wick-Dyson expansion.  Such vacuum-expectation values are routinely
disregarded in purely perturbative calculations, in which the vacuum
is fully annihilated by Fock-space annihilation operators, but must be
taken into consideration if the vacuum has nonperturbative content
(Pascual and Tarrach, 1984).  The configuration-space nonperturbative
propagator necessarily entails a replacement of Eq. (2.4) with the
following expression (Bagan, {\it et al.,} 1994; see also Elias,
{\it et al.,} 1988 and Yndurain, 1989):
\begin{equation}
<0|: \psi (x) \bar{\psi} (y):|0> = \int d^4 k \; e^{-i k \cdot (x-y)}
(\gamma \cdot k+m) {\cal F} (k),
\end{equation}
where
\begin{equation}
\int d^4 k {\cal F} (k) e^{-i k \cdot x} \equiv -<\bar{f}f>J_1 (m
\sqrt{x^2}) / (6m^2 \sqrt{x^2}),
\end{equation}
with $<\bar{f}f>$ identified as the (appropriately normalized) fermion-antifermion
condensate of (1.1).  As is evident from a comparison of (3.1a) to (2.4), the
momentum-space expressions for the nonperturbative propagator
contributions to Fig. 3a and Fig. 3b respectively entail the following 
alterations (Bagan, {\it et al.,} 1994) in the final line of (2.6):

\renewcommand{\theequation}{3.\arabic{equation}}
\setcounter{equation}{1}

\begin{equation}
Fig(3a): \; \; \; \; \; \; \; \; \; \; \frac{1}{(k_2 - p_1)^2 - m^2} \rightarrow
-i (2 \pi)^4 {\cal F} (k_2 - p_1);
\end{equation}
\begin{equation}
Fig(3b): \; \; \; \; \; \; \; \; \; \; \frac{1}{(k_2 - p_2)^2 - m^2}
\rightarrow -i (2 \pi)^4 {\cal F} (k_2 - p_2).
\end{equation}
[The contribution vanishes from the graph in which both fermion
internal-lines are simultaneously altered.]
The net effect of these changes is to reproduce the Feynman amplitude
given in (2.12), but with the Feynman integrals (2.13) altered as
follows:
\begin{eqnarray}
& & \int \frac{d^n k}{(2 \pi)^n}  \frac{[1; \; k_\rho ; \; k_\rho
k_\sigma]}{(k^2-\epsilon^2) [(k-p_2)^2 - m^2][(k-p_1)^2 - m^2]}
\nonumber \\
& & \rightarrow  -i \int \frac{d^4 k [1; \; k_\rho; \; k_\rho
k_\sigma] {\cal F} (k - p_1)}{(k^2 - \epsilon^2) [(k-p_2)^2 - m^2]}
\nonumber \\
& & -i \int \frac{d^4 k [1; \; k_\rho ; \; k_\rho k_\sigma] {\cal F}
(k-p_2)}{(k^2 - \epsilon^2)[(k-p_1)^2 - m^2]}
\end{eqnarray} 
We have returned to explicit use of 4-dimensional integration because
the new integrals are all UV-finite.

Thus we retain the form of the amplitude (2.12), but with the
integrals $I, I_\rho, I_{\rho\sigma}$ now being given (after a
trivial shift of integration variable) by

\begin{eqnarray}
I(p_2, \; p_1) & = & - i \int \frac{d^4 k \; {\cal F}(k)}{[(k+p_1)^2 -
\epsilon^2][(k+p_1 - p_2)^2 - m^2]} \nonumber \\
& & -i \int \frac{d^4 k \; {\cal F} (k)}{[(k+p_2)^2 -
\epsilon^2][(k+p_2 - p_1)^2 - m^2]},
\end{eqnarray}
\begin{eqnarray}
I_\rho (p_2, \; p_1) & = & - i \int \frac{d^4 k (k_\rho + p_{1 \rho})
{\cal F} (k)}{[(k+p_1)^2 - \epsilon^2][(k+p_1 - p_2)^2 - m^2]}
\nonumber \\
& & - i \int \frac{d^4 k (k_\rho + p_{2 \rho}) {\cal F}
(k)}{[(k+p_2)^2 - \epsilon^2][(k + p_2 - p_1)^2 - m^2]} ,
\end{eqnarray}
\begin{eqnarray}
I_{\rho \sigma} (p_2, \; p_1) & = & -i \int \frac{d^4 k (k_\rho +
p_{1\rho})(k_\sigma + p_{1\sigma}) {\cal F} (k)}{[(k+p_1)^2 -
\epsilon^2][(k+p_1 - p_2)^2 - m^2]} \nonumber \\
& & - i \int \frac{d^4 k (k_\rho + p_{2\rho})(k_\sigma + p_{2\rho})
{\cal F} (k)}{[(k+p_2)^2 - \epsilon^2][(k+p_2 - p_1)^2 - m^2]} . 
\end{eqnarray}
Using parametrizations analogous to (2.14-16), 
\begin{equation}
I(p_2, \; p_1) = i {\cal A} (q^2) / m^2,
\end{equation}
\begin{equation}
I_\rho (p_2, \; p_1) = i [(p_{1\rho} + p_{2\rho})/m^2] {\cal B}
(q^2),
\end{equation}
\begin{eqnarray}
I_{\rho\sigma} (p_2, \; p_1) & = & i g_{\rho\sigma} {\cal C} (q^2)
\nonumber \\
& + & i [(p_{1\rho} p_{1\sigma} + p_{2\rho} p_{2\sigma})/m^2] {\cal
D} (q^2) \nonumber \\
& + & i [(p_{1\rho} p_{2\sigma} + p_{2\rho} p_{1\sigma})/m^2] {\cal
E} (q^2),
\end{eqnarray}
we proceed analogously to the derivation of (2.18) and find that
the $<\bar{f}f>$ contribution to $S(q^2)$ in (2.7) is now given to 
one-loop order by
\begin{equation}
\Delta S(q^2) = 2 {\cal B} (q^2) - 2 {\cal D} (q^2) - 2 {\cal E}(q^2).
\end{equation}
As in (2.24), the fermion-antifermion condensate contribution to the 
anomalous magnetic moment is then found to be 
\begin{equation}
\Delta {\cal K} F_2(0) = -8 \left( e^2 Q^2 \right)
\left( {\cal B} (0) - {\cal D} (0) - {\cal E} (0) \right).
\end{equation}

To proceed further, we need to evaluate the integrals (3.6) and (3.7)
that determine the explicit functions ${\cal B}(q^2)$, ${\cal
D}(q^2)$ and ${\cal E}(q^2)$ of Eqs. (3.9) and (3.10). To evaluate 
${\cal B}(q^2)$, we need to evaluate the integrals
$I_\rho (p_2, \; p_1)$ in Eq. (3.6). Using a Feynman-parameter
combination of the propagator denominators, we find for on-shell
momenta $(p_1^2 = p_2^2 = m^2)$ that
\begin{eqnarray}
I_\rho (p_2, \; p_1) & = & -i \int_0^1 dz\int\frac{d^4 k (k_\rho
+ p_{1\rho}) {\cal F}(k)}{\left\{[k - (p_2 z - p_1)]^2 - m^2 z^2 -
\epsilon^2 (1-z) \right\}^2} \nonumber \\
& -i & \int_0^1 dz\int\frac{d^4 k (k_\rho + p_{2\rho}) {\cal F}(k)}{
\left\{ [k - (p_1 z - p_2)]^2 - m^2 z^2 - \epsilon^2 (1 - z)
\right\}^2}
\end{eqnarray} 
We will drop the photon mass $\epsilon^2$, as any infrared divergences 
that might arise {\it in
the anomalous magnetic moment} (as opposed to $F_1(q^2)$) cannot be 
removed by bremsstrahlung
corrections.  The integrals in (3.13) can be expressed in terms of
the integrals (A.6,7) of the Appendix, with $p \equiv p_1 z - p_2$ or
$p_2 z - p_1$, and with $\mu = mz$.  For on-shell momenta, both
definitions of $p$ lead to $p^2 = m^2(1-z)^2 + q^2 z$ with $q^2 =
(p_2 - p_1)^2$. The results we obtain are valid only for $q^2 > 0$;  
the requirement that $p^2 > 0$, as discussed
immediately following (A.9), necessarily implies $q^2 > 0$, as $z$
in $p^2 = m^2(1-z)^2 + q^2 z$ ranges over values between zero and one.  
We then see from (3.13) that
\begin{eqnarray}
I_\rho (p_2, \; p_1) & = & -i p_{1\rho} \int_0^1 dz R_3 (p_2 z - p_1,
\; mz) \nonumber \\
& & -i p_{2\rho} \int_0^1 dz R_3 (p_1 z - p_2 , \; mz) \nonumber \\
& & -i \int_0^1 dz \int \frac{d^4 k \; k_\rho {\cal F} (k)}{ \left\{ [k
- (p_2 z - p_1)]^2 - m^2 z^2 \right\}^2} \nonumber \\
& & -i \int_0^1 dz \int \frac{d^4 k \; k_\rho {\cal F} (k)}{ \left\{ [k
- (p_1 z - p_2)]^2 - m^2 z^2 \right\}^2}
\end{eqnarray} 
where, from the Appendix to this paper, we define

\renewcommand{\theequation}{3.15\alph{equation}}
\setcounter{equation}{0}

\begin{equation}
R_3 (p, \; \mu) \equiv \int \frac{d^4k \; {\cal F} (k)}{[(k-p)^2 -
\mu^2]^2},
\end{equation}
\begin{equation}
R_2 (p, \; \mu) \equiv \int \frac{d^4k \; {\cal F} (k)}{[(k-p)^2 -
\mu^2]}.
\end{equation}
The remaining integrals in (3.14) are of the form

\renewcommand{\theequation}{3.1\arabic{equation}}
\setcounter{equation}{5}

\begin{equation}
\int \frac{d^4 k \; k^\rho {\cal F} (k)}{[(k-p)^2 - \mu^2]^2} =
A(p^2) p^\rho.
\end{equation}
If we contract $p_\rho$ into both sides of (3.16) and use the identities
$p \cdot k = - \frac{1}{2} [(k-p)^2 - \mu^2] + \frac{k^2}{2} +
\frac{p^2}{2} - \frac{\mu^2}{2}$ and $k^2 {\cal F}(k) = m^2 {\cal
F}(k)$ [eq. (A.4)], we find that
\begin{equation}
A(p^2) = -\frac{1}{2p^2} R_2 (p, \; \mu) + \frac{m^2+p^2-\mu^2}
{2p^2} R_3 (p, \; \mu).
\end{equation}
Noting that $\mu^2 = m^2 z^2$, $p^2 = m^2(1-z)^2 + q^2 z$, we then
find that 
\begin{eqnarray}
& & I_\rho (p_2, \; p_1) \nonumber \\
& = & -i p_{1\rho} \int_0^1 dz R_3 (p_2 z - p_1,
\; mz) \nonumber \\
& & -i p_{2\rho} \int_0^1 dz R_3(p_1 z - p_2, \; mz) \nonumber\\
& & -i \int_0^1 dz (p_{2\rho}z - p_{1\rho}) \left[
-\frac{1}{2[m^2(1-z)^2 + q^2 z]} R_2(p_2 z - p_1 , \; mz)
\right. \nonumber \\
& & + \left. \frac{2m^2(1-z) + q^2 z}{2[m^2(1-z)^2 + q^2 z]} R_3 (p_2 z -
p_1, \; mz) \right] \nonumber \\
& & -i \int_0^1 dz (p_{1\rho} z - p_{2\rho}) \left[
-\frac{1}{2[m^2(1-z)^2 + q^2 z]} R_2 (p_1 z - p_2, \; mz) \right.
\nonumber \\
& & + \left. \frac{2m^2(1-z)+ q^2 z}{2[m^2(1-z)^2 + q^2 z]} R_3(p_1 z
- p_2, \; mz) \right].
\end{eqnarray}
We see from (A.6) and (A.7) of the Appendix that $R_2(p, \mu)$ and 
$R_3(p, \mu)$ depend on $p$ only through $p^2$:

\renewcommand{\theequation}{3.19\alph{equation}}
\setcounter{equation}{0}

\begin{equation}
R_2(p_1 z - p_2, \; mz) = R_2 (p_2 z - p_1, \; mz) \equiv R_2[z],
\end{equation}
\begin{equation}
R_3(p_1 z - p_2, \; mz) = R_3 (p_2 z - p_1, \; mz) \equiv R_3[z],
\end{equation}

\renewcommand{\theequation}{3.\arabic{equation}}
\setcounter{equation}{19}

and from (3.17) we find that

\begin{eqnarray}
A[z] & \equiv & A(m^2(1-z)^2 + q^2 z) \nonumber\\
& = & \frac{[2m^2(1-z) + q^2z] R_3[z] - R_2[z]}{2[m^2(1-z)^2 + q^2z]}
\end{eqnarray}
By comparing (3.18) to (3.9), we obtain
\begin{equation}
{\cal B} (q^2) = m^2 \int_0^1 dz \left[ (1-z) A [z] - R_3[z] \right]
\end{equation}
To find ${\cal D}(q^2)$ and ${\cal E}(q^2)$ in (3.10) we combine the
propagator denominators of (3.7) to obtain
\begin{eqnarray}
I_{\rho\sigma} & = & -i \int_0^1 dz \int d^4 k \frac{(k_\rho + p_{1\rho})
(k_\sigma+p_{1\sigma}) {\cal F} (k)}{\left\{[k - (p_2 z - p_1)]^2 -
m^2 z^2 \right\}^2} \nonumber \\
& & -i \int_0^1 dz \int d^4 k \frac{(k_\rho + p_{2\rho})
(k_\sigma+p_{2\sigma}) {\cal F} (k)}{\left\{[k - (p_1 z - p_2)]^2 -
m^2 z^2 \right\}^2} \nonumber \\
& = & + i (p_{1\rho} p_{1\sigma} + p_{2\rho} p_{2\sigma}) \int_0^1 dz
\left[ 2A [z] - R_3 [z] \right] \nonumber \\
& & - i (p_{1\rho} p_{2\sigma} + p_{2\rho} p_{1\sigma}) \int_0^1 dz
\left[ 2z A[z] \right] \nonumber \\
& & -i \int_0^1 \frac{dz \; k_\rho k_\sigma {\cal F}(k)}{\left\{ \left[
k - (p_2 z - p_1) \right]^2 - m^2 z^2 \right\}^2} \nonumber \\
& & -i \int_0^1 \frac{dz \; k_\rho k_\sigma {\cal F}(k)}{\left\{ \left[
k - (p_1 z - p_2) \right]^2 - m^2 z^2 \right\}^2} ,
\end{eqnarray}
The remaining integrals in (3.22) are of the form
\begin{equation}
\int \frac{d^4 k \; k_\rho k_\sigma {\cal F}(k)}{\left\{ (k - p)^2 -
\mu^2 \right\}^2} = X(p^2) g_{\rho\sigma} + Y(p^2) p_\rho p_\sigma ,
\end{equation}
in which case we see from comparison of (3.22) to (3.10) that
\begin{equation}
{\cal D} (q^2) = m^2 \int_0^1 dz \left[ 2A[z] - R_3 [z] - (1+z^2)
Y[z] \right],
\end{equation}
\begin{equation}
{\cal E} (q^2) = +m^2 \int_0^1 dz \left[ -2zA[z] + 2z Y[z] \right],
\end{equation}
where
\begin{equation}
Y[z] \equiv Y(m^2(1-z)^2 + q^2 z).
\end{equation}
Thus we see through comparison of (3.11) to (3.21), (3.24) and (3.25)
that
\begin{equation}
\Delta S(q^2) = 2m^2 \int_0^1 dz \left[ (1-z)^2 Y[z] - (1-z)A[z] \right].
\end{equation}
To determine $Y[z]$, we first note from (A.4) in the Appendix that contraction 
of $g^{\rho\sigma}$ into (3.23) yields the relation
\begin{equation}
m^2 R_3 (p, \; \mu) = 4 X (p^2) + p^2 Y(p^2).
\end{equation}
Recalling that $p \cdot k = -\frac{1}{2} \left[ (k - p)^2 - \mu^2
\right] + \frac{1}{2} (m^2 + p^2 - \mu^2)$, we see that contraction
of $p^\rho p^\sigma$ into (3.23) yields the relation
\begin{eqnarray}
\frac{1}{4} R_1 & - & \frac{1}{2} (m^2 + p^2 - \mu^2) R_2 (p, \; \mu)
+ \frac{1}{4} (m^2 + p^2 - \mu^2)^2 R_3 (p, \; \mu) \nonumber \\
& = & p^2 X(p^2) + (p^2)^2 Y(p^2),
\end{eqnarray}
where $R_1 = \int d^4 k {\cal F} (k)$ = $-<\bar{f}f>/12m$, as shown
in the Appendix.
Given $\mu = mz$, $p^2 = m^2(1-z)^2 + q^2 z$, one can then solve (3.28)
and
(3.29) for $Y(p^2)$ to obtain the following:
\begin{eqnarray}
Y[z] & = & -\frac{1}{3[m^2(1-z)^2 + q^2z]} \left\{ m^2 R_3 [z]
\right. \nonumber \\
& - & \frac{1}{(m^2(1-z)^2 + q^2z)} \left[ R_1 - 2(2m^2(1-z) + q^2 z)
R_2[z] \right. \nonumber \\
& + & \left. \left. [2m^2 (1-z) + q^2 z]^2 R_3 [z] \right] \right\}.
\end{eqnarray}
The integral (3.27) can be evaluated using the expressions (3.20) and
(3.30) to express $A[z]$ and $Y[z]$ in terms of $R_1$, $R_2[z]$ and
$R_3[z]$.  We note from (A.6) and (A.7) of the Appendix that for 
$p^2 = m^2 (1-z)^2 +
q^2 z$, $\mu^2 = m^2 z^2$, $4m^2 > q^2 > 0$, that
\begin{equation}
R_2[z] = \frac{<\bar{f}f>}{24m^3} \frac{[-2m^2(1-z) - q^2z + iz
\sqrt{4m^2q^2-q^4}]}{m^2(1-z)^2 + q^2 z},
\end{equation}
\begin{equation}
R_3[z] = \frac{<\bar{f}f>}{24m^3[m^2(1-z)^2 + q^2 z]} \left[ 1 +
i \frac{[2m^2(1-z)+q^2z]}{z \sqrt{4m^2 q^2-q^4}} \right].
\end{equation}
We note that $R_2$ and $R_3$ have developed imaginary parts when
$q^2$ is between zero and $4m^2$; $R_2$ and $R_3$ are both real if
$q^2 > 4m^2$.  Such a branch cut between $q^2 = 0$ and $q^2 = 4m^2$
is also seen to occur in the quark-antiquark condensate contributions
to two-point current-correlation functions (Bagan {\it et al.,} 1986;
Elias {\it et al.,} 1993), and is discussed in detail in the section
that follows.  We find from substitution of (3.20) and (3.30-3.32)
into (3.27) that 
\begin{eqnarray}
\Delta S(q^2) & = & 2m^2 \int_0^1 dz (1-z) \left[ -\frac{<\bar{f}f>
(1-z)}{36m[m^2(1-z)^2 + q^2 z]^2} \right. \nonumber \\
& - & \frac{[5m^2(1-z)^2 + q^2 z (1-4z)]}{6[m^2(1-z)^2 + q^2
z]^2} R_2[z] \nonumber \\
& - & \left. \frac{[m^2 q^2 z(1-z)(3+5z) + q^4 z^2
(1+2z)]}{6[m^2(1-z)^2 + q^2 z]^2} R_3 [z] \right]. 
\end{eqnarray}
Explicit evaluation of the real and imaginary parts of (3.33) for $0
< q^2 < 4m^2$ yields the following results:
\begin{equation}
Re [\Delta S (q^2)] = 0 ,
\end{equation}
\begin{equation}
Im [\Delta S (q^2)] = -\frac{<\bar{f}f>}{12m \sqrt{4m^2 q^2 - q^4}}
\end{equation}
where $\Delta {\cal K} F_2 (q^2) = - 4 e^2 Q^2 \Delta S (q^2)$.

\bigskip

\noindent{\bf 4. DISCUSSION}

\noindent{\bf 4.1 Gauge Invariance}

In an arbitrary covariant gauge, the contribution of Fig 3a is
proportional to

\renewcommand{\theequation}{4.\arabic{equation}}
\setcounter{equation}{0}

\begin{eqnarray}
& & \int d^4 k \;  {\cal D}^{\tau\sigma} (k, \xi) \frac{{\cal F} (k - p_1)
\gamma_\tau (\not{k} - \not{p}_2 - m) \gamma_\mu (\not{k} - \not{p}_1
- m) \gamma_\sigma}{[(k - p_2)^2 - m^2]} \nonumber \\
& & \equiv  \Lambda^{(a)} (p_2, \; p_1) - (1 - \xi) \Lambda_\xi^{(a)}
(p_2, \; p_1),
\end{eqnarray}
as is evident from (3.3) and (2.6), where $\xi$ is the
photon-propagator gauge parameter
\begin{equation}
{\cal D}^{\tau\sigma} (k) = \frac{g^{\tau\sigma}}{k^2} - (1 - \xi)
\frac{k^\tau k^\sigma}{k^4} .
\end{equation}
In (4.1), $\Lambda^{(a)}$ is the (Feynman-gauge) contribution we have
already considered, and $\Lambda_\xi^{(a)}$ is the contribution
arising from the second term in (4.2). Gauge-parameter independence
is explicit provided $\bar{u}(p_2) \Lambda_\xi^{(a)} (p_2, \; p_1)
u(p_1) = 0$; i.e., provided the $k^\tau k^\sigma / k^4$ term in
${\cal D}^{\tau\sigma}$ does not contribute to the on-shell vertex
correction. To demonstrate this, we consider
\begin{eqnarray}
& & \Lambda_\xi^a (p_2, \; p_1) u (p_1) \nonumber \\
&  = & \int \frac{d^4 k \; {\cal F}(k - p_1) \not{k} (\not {k} -
\not{p}_2 - m) \gamma_\mu (\not{k} - \not{p}_1 - m) \not{k}}{k^4 [(k
- p_2)^2 - m^2]} u(p_1) \nonumber \\
& \begin{array} {c}{} \\ \Longrightarrow \\ _{k-p_1 \rightarrow
k}
\end{array} &
\int \frac{d^4 k \; (\not{k} + \not{p}_1)(\not{k} + \not{p}_1 -
\not{p}_2 - m) \gamma_\mu {\cal F}(k) (\not{k} -
m)(\not{k}+\not{p}_1) u(p_1)}{\left( (k+p_1)^2 \right)^2 \left[
(k+p_1 - p_2)^2 - m^2\right]} \nonumber \\
\end{eqnarray}
${\cal F}(k)$ is a Dirac scalar that can be moved past gamma-matrices --
we have moved it to the right in (4.3) in order to focus on the
factors immediately preceding $u(p_1)$:
\begin{eqnarray}
& & {\cal F}(k) (\not{k} - m) (\not{k} + \not{p}_1) u (p_1) \nonumber \\
& & = {\cal F} (k) (k^2 - m^2) u(p_1) \nonumber \\
& & = 0.
\end{eqnarray}
The second to last line of (4.4) is a consequence of
${\not p}_1 u
(p_1) = mu(p_1)$, and the final line is a consequence of $k^2 {\cal
F} (k) = m^2 {\cal F} (k)$, as discussed in the Appendix [eq. (A.4)]. 
Thus we see that $\Lambda_\xi^a (p_2, \; p_1) u(p_1) = 0$. A
virtually identical argument shows the gauge-dependent contribution
from Fig. 3b annihilates $\bar{u}(p_2)$ on-shell.  Consequently, we
see that the vertex correction $\bar{u} (p_2) \Lambda^\mu (p_2, \;
p_1) u(p_1)$ is manifestly gauge-parameter independent: \\
$\bar{u}(p_2) \Lambda_\xi^{(a)} (p_2, \; p_1) u (p_1) = 0$.  

This demonstration of gauge-parameter independence, however, is
contingent upon having {\it the same} fermion mass enter the
perturbative and nonperturbative  fermion propagators (2.4) and
(3.1), as has been assumed throughout the previous section's
calculation.  Any attempt to distinguish between these masses will
destroy the gauge-parameter independence of the result ({\it
e.g.,} He, 1996). The gauge-parameter independence of electroweak
two-point functions has similarly been shown (Ahmady, {\it et al.,}
1989) to be contingent, for a given flavour, upon having the same
fermion mass enter from nonperturbative vacuum expectation values
(1.1) as appears in the corresponding fermion propagator function
(1.2).

An entirely analogous situation arises in QCD when one considers the
fermion-antifermion condensate contribution to the fermion two-point
function.  The apparent gauge-parameter dependence first seen for
this contribution (Pascual and de Rafael, 1982) has been shown to
disappear on-shell (Elias and Scadron, 1984) provided the mass that
appears in the fermion propagator (2.4) is consistent with that
appearing in the vacuum expectation value (1.1).  Since this
latter mass is necessarily dynamical, gauge-parameter independence
suggests that the fermion mass appearing
throughout the calculation of the previous section be understood to
be dynamical rather than Lagrangian in origin (Elias and Scadron,
1984; Reinders and Stam, 1986), a reflection of the chiral
noninvariance of the vacuum necessary for (1.1) to be nonzero ({\it
i.e.,}
for $<\bar{f}f> \neq 0$).

\newpage

\noindent{\bf 4.2 Interpretation of ${\bf Im (\Delta S (q^2))}$}

As noted in the previous section, the integrals $R_2[z]$ and
$R_3[z]$ are seen to contribute an imaginary part to the vertex
function when $q^2$ is between $0$ and $4m^2$, behaviour that is also evident
in the fermion-antifermion condensate contributions to two-point
functions (Bagan, 1986; Elias {\it et al.,} 1993).  Although
imaginary parts of Feynman amplitudes are signals of physical
intermediate states, the region $0 < q^2 < 4m^2$ is clearly beneath
the $q^2 = 4m^2$ kinematic threshold for the production of a physical
fermion-antifermion $(\bar{f}f)$ pair.  Nevertheless, this $0 < q^2 <
4m^2$ branch cut, when augmented by the purely-perturbative
$\bar{f}f$-production branch cut beginning at $q^2 = 4m^2$, may be
associated with the $q^2 = 0$ production threshold for Goldstone
bosons associated with chiral symmetry breaking.  The internal
consistency of such a picture necessitates identification of $m$ with
a dynamical mass, as opposed to a mass that appears in the Lagrangian
and that explicitly breaks Lagrangian chiral symmetry; {\it e.g.} the pion
is massless only in the limit of Lagrangian chiral symmetry (zero
current-quark mass).  As noted above, such a dynamical fermion mass
is expected to arise from the chiral-noninvariance of the QCD vacuum
itself, and can be related directly to the $<\bar{f}f>$
order-parameter characterizing chiral non-invariance (Politzer, 1976;
Elias and Scadron, 1984).  We have also seen above that such an
interpretation is strongly suggested by gauge invariance.  Thus it
would appear that the nonzero imaginary part occurring in (3.35) may
be a kinematic manifestation of the Goldstone theorem, suggesting
that the theory now contains the zero-mass meson anticipated from a
{\it dynamical} breakdown ($<\bar{f}f> \neq 0$) of Lagrangian chiral
symmetry.

\bigskip

\noindent{\bf 4.3 Quarks?}

Although the field-theoretical calculations presented in this paper 
have been posed almost entirely in the abstract, there are clear reasons 
to explore
their applicability to the electromagnetic properties of quarks. 
Quarks couple to both QED and QCD interactions.  Even though the
latter are deemed entirely responsible for the existence of
$<\bar{q}q>$ condensates, such condensates necessarily contribute to
Feynman amplitudes from which quark electromagnetic properties are
extracted.

As remarked in the Introduction, such properties are of evident
interest to quark-model estimates of hadron properties.  
For example, the construction of proton and neutron magnetic moments from
the magnetic moments of Dirac-fermion quark constituents (Beg. {\it
et al.,} 1964), one of the very earliest successes of the nonrelativistic 
quark model, is sensible only if the quark masses employed are
vastly larger (${\cal O}$(300 MeV)) than those 
masses anticipated from the QCD Lagrangian (${\cal O}$(5-10 MeV)).
Thus, there is a clear 
phenomenological role for a dynamical mass in quark-model physics,
even though such a larger mass may really represent the inverse
length of a confinement radius.

Any application of the calculation presented in Section 3 to quark
electro-magnetic-properties needs to recognize that QED and QCD cannot
be treated in isolation.  Not only is the chiral-noninvariant QCD
vacuum the ``standard-model'' vacuum that quarks actually experience,
suggesting the need to include $<\bar{q}q>$-contributions; 
the photon exchanges of QED in
isolation must also be augmented by gluon exchanges. Consequently,
the spin-1 internal lines of Fig 3 represent photons {\it and}
gluons, entailing substitution of [$e^2 Q^2 + g_s^2 T(R)$] for
all factors of $e^2 Q^2$ appearing in Section 3.  This change becomes
problematical in the $q^2 \rightarrow 0$ limit, a region believed to
be inaccessible to perturbative QCD, though there exists evidence
(and considerable prejudice) for a freezing out of the effective QCD
coupling $g_s$ to not-overly large values at small momentum transfers
(Mattingly and Stevenson, 1992; Stevenson, 1994; Ellis {\it et al.,}
1997; Baboukhadia {\it et al.,} 1997; Gardi and Karliner, 1998).  
Similarly, the nonabelian character
of QCD leads to a vastly richer set of gluon bremsstrahlung graphs
contributing to $F_1(q^2)$;  it is precisely for this reason we have
focused on ${\cal K} F_2$, a quantity insensitive to such graphs.

Subject to all these concerns, it is of interest to speculate on the
applicability of the previous section's results to phenomenological
quark properties. Naively, the result (3.34) would imply the absence
of a $<\bar{q}q>$ contribution to the quark magnetic moment,
particularly if the divergence in the imaginary part at $q^2 = 0$
(3.35) is attributable to the production of zero-mass pions.  {\it
Moreover, an alteration in the kinematic production threshold from $4m^2$
to zero may reflect QCD's transition from a quark-gluon gauge theory
to true low-energy hadronic physics}.  As discussed above, $m$ is
understood in such a picture to be the ${\cal O}$(300 MeV) dynamical quark
mass characterizing the quark magneton in the static
quark model.  The success of such a picture appears not to be
compromised by $<\bar{q}q>$ effects, if (3.34) is taken at
face-value.

When $q^2 > 4m^2$, however, the quark-condensate
contribution to ${\cal K} F_2(q^2)$ is entirely real:

\begin{equation}
\Delta S (q^2) = \frac{ < \bar{q}q>}{12m \sqrt{q^4 - 4m^2 q^2}}.
\end{equation}
Even though $m$ is understood here to be dynamical [$<\bar{q}q> / m
\sim m^2$(Politzer, 1976; Elias and Scadron, 1984)], 
we have no explanation for the divergence in (4.5)
as $q^2 \rightarrow 4m^2$ from above.  One does see from (4.5) that
the quark-condensate contribution to ${\cal K} F_2 (q^2)$ goes like
$1/q^2$ in the large $q^2$-limit.  Such behaviour also characterizes
(up to logarithms) the purely-perturbative contributions to ${\cal K}
F_2 (q^2)$ discussed in Section 2, and can be linked via
quark-counting rules to the $1/Q^6$ behaviour of the nucleon form
factor $F_2^N (Q^2)$ (Brodsky and Lepage, 1989).

Finally, we point out that the {\it large-$Q^2$} behaviour of the
quark-condensate contribution (4.5) is not suppressed relative to the
purely-perturbative contribution.  On dimensional grounds, one might
expect an ${\cal O}[m<\bar{q}q> / Q^4]$ quark-condensate contribution
that is suppressed relative to the purely perturbative contribution
at short distances.  Such behaviour is clearly unsupported by (4.5), 
which suggests that the nonperturbative order parameter $<\bar{q}q>$ 
may leave footprints even in the deep-inelastic domain.

\bigskip

\noindent{\bf ACKNOWLEDGEMENTS}

We are grateful for discussions with D.G.C. McKeon, V. A. Miransky
and T. G. Steele, and for support from the Natural Sciences and
Engineering Research Council of Canada.

\newpage

\noindent{\bf References}

\begin{description}

\item{}Ahmady, M.R., Elias, V., Mendel, R. R., Scadron, M.D., Steele,
T. (1989) {\it Physical Review D: Particles and Fields}, {\bf 39},
2764.

\item{}Baboukhadia, L. R., Elias, V., and Scadron, M.D. (1997) {\it
Journal of Physics G: Nuclear and Particle Physics}, {\bf 23}, 1065.
\item{}Bagan, E., LaTorre, J. I., and Pascual, P. (1986) {\it Zeitschrift
f\"ur Physik C - Particles and Fields}, {\bf 32}, 43.

\item{}Bagan, E., Ahmady, M. R., Elias, V., and Steele, T. G. (1993)
{\it Physics Letters B}, {\bf 305}, 151.

\item{}Bagan, E., Ahmady, M. R., Elias, V., and Steele, T. G. (1994)
{\it Zeitschrift f\"ur Physik C - Particles and Fields}, {\bf 61}, 157.

\item{}Beg, M. A. B., Lee, B. W., and Pais, A. (1964) {\it Physical 
Review Letters}, {\bf 13}, 514 and 650.

\item{}Bloch, F. and Nordsieck, A. (1937) {\it Physical Review}, {\bf
37}, 54.

\item{}Brodsky, S. J., and Lepage, G. P. (1989) Exclusive processes in 
quantum chromodynamics, in {\it Perturbative Quantum Chromodynamics}, 
A. H. Mueller, ed., World Scientific, Singapore, pp. 153-156.

\item{}Elias, V. and Scadron, M. D. (1984) {\it Physical Review D: 
Particles and Fields}, {\bf 30}, 647.

\item{}Elias, V., Steele, T. G., and Scadron, M. D. (1988) {\it Physical 
Review D: Particles and Fields}, {\bf 38}, 1584.

\item{}Elias, V., Murison, J. L., Scadron, M. D., and Steele, T. G. (1993) 
{\it Zeitschrift f\"ur Physik C - Particles and Fields}, {\bf 60}, 235.

\item{}Ellis, J., Karliner, M., and Samuel, M. A. (1997) 
{\it Physics Letters B}, {\bf 400}, 176.

\item{}Feynman, R. P. (1949) {\it Physical Review}, {\bf 76}, 749 and 769.

\item{}Gardi, E. and Karliner, M. (1998) Preprint [hep-ph/9802218].

\item{}Gradshteyn, I. S. and Ryzhik, I. M. (1980) {\it Table of Integrals, 
Series, and Products}, Academic Press, Orlando, p. 712 [eq. (6.623.3)].

\item{}He, H. (1996) {\it Zeitschrift f\"ur Physik C - Particles and
Fields}, {\bf 69}, 287.

\item{}Higashijima, K. (1984) {\it Physical Review D: Particles and
Fields}, {\bf 29}, 1228.

\item{}Mattingly, A. C., and Stevenson, P. M. (1992) {\it Physical Review 
Letters}, {\bf 69}, 1320.

\item{}Pascual, P. and de Rafael, E. (1982) {\it Zeitschrift f\"ur Physik C 
- Particles and Fields}, {\bf 12}, 127.

\item{}Pascual, P. and Tarrach, R. (1984) {\it QCD: Renormalization for the 
Practitioner}, Lecture Notes in Physics 194, Araki, H., Ehlers, J., Hepp, K., 
Kippenhahn, R., Weidenm\"uller, H. A., and Zittartz, J., eds., Springer-Verlag, 
Berlin, pp. 168-191.

\item{}Perkins, D. H. (1987) {\it Introduction to High Energy
Physics}, Addison-Wesley, Reading, Mass.,
pp. 164-167. 

\item{}Politzer, H. D. (1976) {\it Nuclear Physics B: Particle
Physics}, {\bf 117}, 397.

\item{}Reinders, L. J. and Stam, K. (1986) {\it Physics Letters B},
{\bf 180}, 125. 

\item{}Schwinger, J. (1948) {\it Physical Review}, {\bf 73}, 416.

\item{}Shifman, M. A., Vainshtein, A. I., and Zakharov, V. I. (1979) 
{\it Nuclear Physics B: Particle Physics}, {\bf 147}, 385 and 448.

\item{}Stevenson, P. M. (1994) {\it Physics Letters B}, {\bf 331}, 187.

\item{}Yennie, D. R., Frautschi, S. C., and Suura, H. (1955) {\it Annals of 
Physics}, {\bf 13}, 379.

\item{}Yndurain, F. J. (1989) {\it Zeitschrift f\"ur Physik C - Particles 
and Fields}, {\bf 42}, 643.

\end{description}

\newpage 

\noindent{\bf APPENDIX:  Feynman Integrals over the Nonperturbative
Propagator Function ${\cal F} (k)$}

\renewcommand{\theequation}{A.\arabic{equation}}
\setcounter{equation}{0}

The vacuum expectation value of a normal ordered pair of
condensing fermion-antifermion fields may be expressed as
follows (Bagan, {\it et al.,} 1993 and 1994; Yndurain, 1989):

\begin{eqnarray}
& & <0|:f(x)\bar{f}(0):|0> \nonumber \\
& = & -\frac{<\bar{f}f>}{6m^2} \left( i
\gamma^\mu \partial_\mu + m \right) \left[ J_1 \left( m \sqrt{x^2}
\right) / \sqrt{x^2} \right] \nonumber \\
& \equiv & \int d^4 k \; e^{-ikx} \left( \gamma^\mu k_\mu + m \right)
{\cal F} (k),
\end{eqnarray}
where ${\cal F}$ is the Fourier transform
\begin{equation}
\int d^4 k \; e^{-ik \cdot x} {\cal F} (k) = -\frac{<\bar{f}f>}{6m^2}
\frac{J_1 \left( m \sqrt{x^2}\right)}{\sqrt{x^2}}
\end{equation}
which is well-defined for causal Minkowskian separations $(x^2 > 0)$. 
The normalization chosen for $<\bar{f}f>$ is discussed in Section 1.
The second line of (A.1) is a solution to the free-particle Dirac
equation,
\begin{equation}
\left( i \gamma^\mu \partial_\mu - m \right) <0|:f(x) \bar{f}(0):|0>
= 0.
\end{equation}
Application of (A.3) to the final expression on the right-hand side of 
(A.1) implies that (Bagan, {\it et al.,} 1994)
\begin{equation}
k^2 {\cal F}(k) = m^2 {\cal F}(k).
\end{equation}

Let us now consider the following integrals arising in the text:

\begin{equation}
R_1 \equiv \int d^4 k \; {\cal F} (k) = -<\bar{f}f> / 12m
\end{equation}  

\begin{eqnarray}
& & R_2 (p, \; \mu) \equiv \int \frac{d^4 k \; {\cal F}(k)}{(p-k)^2 -
\mu^2 + i |\epsilon|} = - \frac{<\bar{f}f>}{24m^3 p^2} \nonumber \\ 
& \times & \left[ p^2 + m^2 - \mu^2 -
\sqrt{[p^2 -(m-\mu)^2][p^2 - (m+\mu)^2]} \right], 
\end{eqnarray}

\begin{eqnarray}
R_3 (p, \; \mu) & & \equiv \int \frac{d^4 k \; {\cal F}(k)}{\left[ (p-k)^2 -
\mu^2 + i |\epsilon| \right]^2} \nonumber \\ 
& = & \frac{<\bar{f}f>}{24m^3 p^2} \left[ 1 - \frac{p^2 + m^2 - \mu^2}{[p^2 
- (m - \mu)^2][p^2 - (m+\mu)^2]} \right]. 
\end{eqnarray} 
Eqs. (A.6) and (A.7) are demonstrably valid only for $p^2 > 0$, as
discussed below.

The integral $R_1$ is obtained from the $x \rightarrow 0$ limit of
(A.2).  To evaluate the integral $R_2(p, \; \mu)$ we first utilize
(A.4) to replace $k^2$ with $m^2$, and we then utilize the positivity of
$|\epsilon|$ to exponentiate the propagator:
\begin{eqnarray}
R_2(p, \; \mu) = -i \int_0^\infty d\eta \; e^{i\eta(p^2 + m^2 - \mu^2 +
i |\epsilon|)} \int d^4 k \; e^{-ik \cdot(2p\eta)} {\cal F}(k)
\nonumber \\
= \frac{i <\bar{f}f>}{6m^2} \int_0^\infty d\eta \; e^{-\eta[|\epsilon| -
i(p^2 + m^2 - \mu^2)]} J_1 \left( 2\eta m \sqrt{p^2} \right) / \left(
2\eta \sqrt{p^2} \right),
\end{eqnarray}
where the final line of (A.8) is obtained directly from (A.2) with
$x$ replaced by $2\eta p$.  The resulting integral over $\eta$ is
evaluated through use of the tabulated integral (Gradshteyn and
Ryzhik, 1980)
\begin{equation}
\int_0^\infty e^{-\eta\alpha} J_1(\eta \beta) d\eta / \eta =
\frac{\sqrt{\alpha^2 + \beta^2} - \alpha}{\beta} \; , \; (Re \; \alpha >
|Im \; \beta|).
\end{equation}

Since $Re \; \alpha$ is identified with the positive infinitesimal 
$|\epsilon|$ and $\beta$ is
identified with $2m\sqrt{p^2}$ in (A.8), $p^2$ must be positive.
The following results are obtained for physical (Minkowskian)
momenta:

\begin{equation}
R_2(p, \; \mu) = \frac{i<\bar{f}f>}{12m^2\sqrt{p^2}} \left[ -i
\frac{ \left[ \sqrt{(p^2+m^2-\mu^2)^2 - 4m^2 p^2} - (p^2 + m^2 -
\mu^2) \right] }{2m\sqrt{p^2}}\right],
\end{equation} 
a result easily rearranged to yield (A.6).

To evaluate the integral $R_3(p, \; \mu)$, we again utilize (A.4) to
replace $k^2$ with $m^2$ and then exponentiate the propagator:
\begin{eqnarray}
R_3(p, \; \mu) = - \int_0^\infty d\eta \; \eta \; e^{i\eta(p^2 + m^2 - \mu^2
+ i|\epsilon|)} \int d^4 k \; {\cal F}(k) e^{-ik \cdot (2\eta p)}
\nonumber \\
= \frac{<\bar{f}f>}{12m^2 \sqrt{p^2}} \int_0^\infty d\eta \;
e^{-\eta[|\epsilon| - i(p^2+m^2-\mu^2)]} J_1 \left( 2 \eta \; m
\sqrt{p^2} \right).
\end{eqnarray}
The final integral in (A.11) is evaluated through use of the integral
\begin{equation}
\int_0^\infty e^{-\eta \alpha} J_1(\eta \beta) d \eta =
\frac{\sqrt{\alpha^2 + \beta^2} - \alpha}{\beta \sqrt{\alpha^2 +
\beta^2}},
\end{equation}
by taking the partial derivative of both sides of (A.9) with respect to 
$\alpha$.  Using (A.12), which is again demonstrably valid 
$(|\epsilon| > |Im 2\eta \sqrt{p^2}|)$
only for $p^2 > 0$, we easily obtain the result (A.7).

It is evident from these procedures that any Feynman integrals of the
form
\begin{equation}
R_{N+1} (p, \; \mu) \equiv \int \frac{d^4 k \; {\cal F}(k)}{[(p-k)^2
- \mu^2 + i|\epsilon|]^N}
\end{equation}
can be evaluated through use of (A.2) and (A.4), exponentiation of the
denominator, and successive differentiations of (A.9):
\begin{eqnarray}
R_{N+1} (p, \; \mu) & = & \frac{(-i)^N}{(N-1)!} \int_0^\infty d\eta
\; \eta^{N-1} e^{-\eta[|\epsilon| - i(p^2 + m^2 - \mu^2)]} \nonumber \\
& \times & \int d^4 k \; {\cal F}(k) e^{-ik \cdot(2p\eta)} \nonumber \\
& = & - \frac{(-i)^N}{(N-1)!} \frac{<\bar{f}f>}{12m^2 \sqrt{p^2}}
\int_0^\infty d\eta \; \eta^{N-2} e^{-\eta[|\epsilon| - i(p^2 + m^2 -
\mu^2)]} J_1 \left( 2m\eta \sqrt{p^2} \right) \nonumber \\
& = & \frac{i^N <\bar{f}f>}{(N-1)! 12m^2 \sqrt{p^2}}
\frac{\partial^{N-1}}{\partial \alpha^{N-1}} \left(
\frac{\sqrt{\alpha^2 + \beta^2} - \alpha}{\beta} \right)
\left|_{\stackrel{\alpha = -i(p^2 + m^2 - \mu^2)}{\beta =
2m\sqrt{p^2}}} \right. .
\end{eqnarray}

\newpage

{\bf Figure Captions:}

\begin{description}
\item[Figure 1:] The fermion-antifermion-photon Green's
function.

\item[Figure 2:] The one-loop purely-perturbative contribution to the 
fermion-antifermion-photon Green's function in configuration space.

\item[Figure 3:] The leading fermion-antifermion condensate contributions to the 
fermion-antifermion-photon Green's function in configuration space, with
                 nonperturbative propagators replacing internal fermion lines as indicated
                 in (3.2) and (3.3).
\end{description}
\end{document}